\documentclass[prb,twocolumn,floatfix,groupedaddress]{revtex4}
\usepackage{graphicx,amsfonts,amssymb,hyperref,nicefrac,lipsum, verbatim}

\usepackage[fleqn]{amsmath}
\usepackage[version=3]{mhchem}
\usepackage{verbatim}
\usepackage{mathtools}

\usepackage{siunitx}  % Einheiten und Zahlenformatierung
\usepackage[x11names]{xcolor}
\newif\ifhyper
\hypertrue
\ifhyper
\hypersetup{
   citecolor = {green!60!black},
   colorlinks = {true}, % false
   urlcolor = {blue} % magenta
}
\fi

\newcommand{\be}{\begin{equation}}
\newcommand{\ee}{\end{equation}}
\newcommand{\beqa}{\begin{eqnarray}}
\newcommand{\eeqa}{\end{eqnarray}}

\def\Longarrow{\protect\@lra}
\def\@lra{\relbar\joinrel\relbar\joinrel\relbar\joinrel%
          \relbar\joinrel\rightarrow}

\begin{document}

\title{Tunable signal velocity in the integer quantum Hall effect of tailored graphene}

\author{M.~Malki}
%\email{maik.malki@tu-dortmund.de}
\affiliation{Lehrstuhl f\"ur Theoretische Physik 1, TU Dortmund, Germany}

\author{G.\ S.~Uhrig}
\affiliation{Lehrstuhl f\"ur Theoretische Physik 1, TU Dortmund, Germany}
\date{\rm\today}

\begin{abstract}
Topological properties in condensed matter physics are often claimed to be a fruitful
resource for technical applications, but so far they only play a minor role in 
applications. Here we propose to put topological edge states 
to use in tailored graphene for Fermi velocity engineering. By tuning external control parameters such as gate voltages, the dispersions of the edge states regime are modified in a controllable way. This enables the realizations of devices such as tunable delay lines and interferometers with switchable delays.
\end{abstract}

\maketitle

\section{Introduction}

Non-trivial topological properties of condensed matter systems are believed to 
represent a valuable resource for various purposes. The key idea is that
topological properties are protected so that they are not destroyed by small
changes of the system. Hence they are robust against imperfections and unwanted effects.
An excellent  example is the quantized Chern number in Chern insulators.
It is well-established that integer quantum Hall systems represent such Chern insulators and
that their Hall conductivity is proportional to the Chern number and thus extremely
well quantized \cite{thoul82,avron83, niu85, kohmo85}. This has led to the most 
spectacular application
of a topological insulator: the integer quantum Hall effect (IQHE) has become the 
international gauge standard for resistance measurements, see Ref.\ \onlinecite{weis11} and 
references therein.

\begin{figure}
	\centering
		\includegraphics[width=0.75\columnwidth]{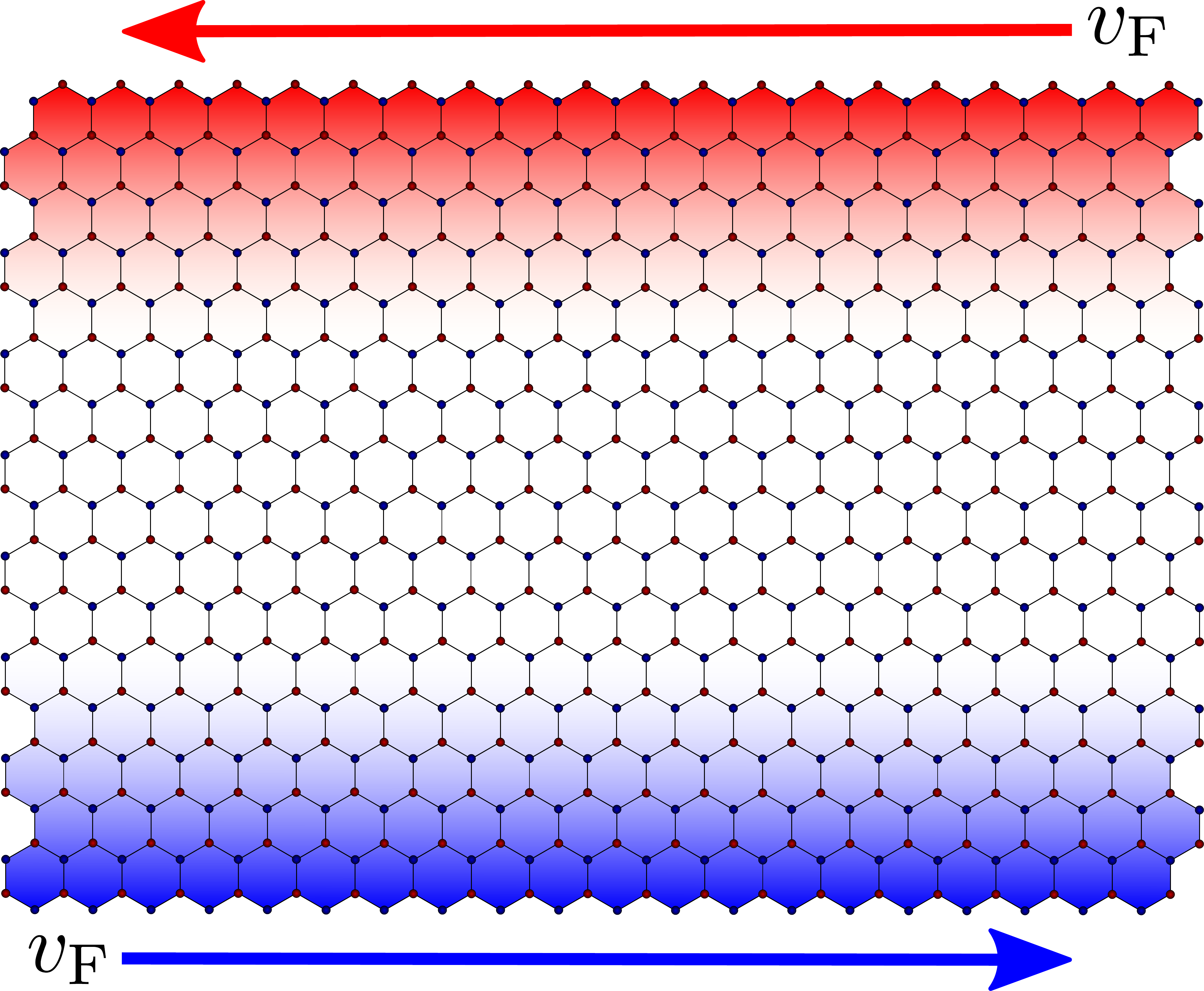}
	\caption{Sketch of a strip of a generic Chern insulator; concretely a strip of graphene. 
	The localization of the chiral edge states are indicated by blue and red shadings whereas the direction are shown by  arrows in the same color.}
	\label{fig:edge_state}
\end{figure}

Apart, however, from this very important application there has been little application
of topological properties so far. We are aware of three-dimensional topological insulators 
used as thermoelectric elements \cite{kadel11, perry11}.
Recently, it has been proposed that the Fermi velocity of electrons in edge states
of two-dimensional Chern insulators on lattices can be tuned by the design of the edges and by
changing the potential of the outermost sites of a strip of the lattice \cite{uhrig16,malki17b}.
It was conjectured that such systems enable one to tune the signal velocity of a charge
signal propagating along the edges by controlling external parameters such as gate voltages.
The exponentially localized edge states possess a chiral nature, i.e., 
the propagation in one direction takes place 
at one edge while propagation in the opposite direction takes place at the other edge, 
see Fig.\ \ref{fig:edge_state} for a generic illustration. Thus tuning of edge-specific 
properties renders the control of velocities depending on direction possible.
The promise is to realize direction-dependent delay lines and interferometers.

The site-specific control of Chern insulators on lattices is a tremendous challenge to
experimental realization. 
Thus, it suggested itself to use the well-established IQHE
for the same purpose. Indeed, it is possible to obtain tunable signal velocities
in two-dimensional electron gases (2DEG) subjected to a perpendicular magnetic field \cite{malki17c}.
Still, there are challenges opposing an immediate realization: modifying the edges by periodically aligned bays with the required precision on small length scales of $\SI{100}{nm}$
represents a tremendous task to sample design. Larger length scales are easier to realize,
but the characteristic length $l^2_\text{B}=h/(e|B|)$ must match the geometric scales so that
larger length scales require smaller magnetic fields. At first sight, this seems easy to 
realize, but the mobilities in the 2DEGs are not high enough to allow for the observation
of the IQHE at low magnetic fields.

For this reason, we advocate to explore alternative routes and it is natural to 
look for other systems displaying an IQHE. Graphene and related compounds are obvious 
candidates. Graphene is widely known for its special electronic properties \cite{zhang05,neto09} and its extraordinary structure. It represents an isolated single sheet of graphite 
\cite{novos05, geim07} and as such realizes a  two-dimensional (2D) allotrope of carbon. 
The low-energy band structure of graphene comprises two Dirac cones distinguished by different locations in the Brillouine zone. Thus, electrons near the Fermi level have a linear dispersion relation and therefore behave like massless relativistic particles. Theoretically, low-energy electrons are described by the Dirac equation \cite{divin84, semen84} where the speed of light is replaced by the Fermi velocity \smash{$v_\mathrm{F} \approx \SI{e6}{\meter/\second}$} \cite{hwang12b, liu15}.
Engineering this important parameter has been realized already by varying the substrate \cite{hwang12b}.

Subjecting graphene to a strong magnetic field at low temperatures leads to the formation of 
relativistic Landau Levels (LL). As a result, one can observe an unconventional IQHE 
\cite{gusyn05, zhang05}. The Hall conductivity in graphene appears 
at half-integer values: $\sigma_{xy} = \pm 4 e^2/h (|n| + 1/2)$. The four-fold degeneracy given 
by valley and spin degeneracy yields the prefactor of $4$. Comparing $\sigma_{xy}(B)$ with the  
IQHE of a non-relativistic 2DEG, the offset $1/2$ can be attributed to the single LL 
with energy $E_0 = 0$ \cite{gusyn05}. This LL is intrinsically half-filled so that one half 
contributes to the valence band and causes the half-integer conductivity. 
Due to cleaner  samples and more precise measuring instruments the IQHE can be detected down to
small external magnetic fields as low as $\approx \SI{0.1}{T}$ due to a very high 
electron mobility \cite{bolot08}.
The unique properties of graphene open up many new possibilities for the basic research 
and technical application, especially in electronics \cite{neto09, dean10} and spintronics 
\cite{han14, roche15}.  

Our central proposal is to use graphene (or a related system) as IQHE system with
accurately tailored edges in order to realize a tunable signal velocity $v_\mathrm{F}$.
We point out that tuning $v_\mathrm{F}$ does not influence the  DC conductivity often studied 
in literature. In our view, there are several advantages of graphene as basis material over 
2DEGs in semiconductors: 
(i) the high mobility allows one to reach the IQHE even at low magnetic fields 
($\approx 0.1$T \cite{bolot08});
(ii) the possibilities to modify the edges in a reproducible and accurate way are larger;
(iii) the energy separation between the lowest LLs ($n < 3$) are much larger
so that the IQHE in graphene can be observed for lower magnetic fields and higher temperatures 
\cite{lafon15}. So we expect that the tunability of the Fermi velocity $v_\mathrm{F}$ 
in the edge modes will be feasible in the near future.

\begin{figure}
	\centering
		\includegraphics[width=0.8\columnwidth]{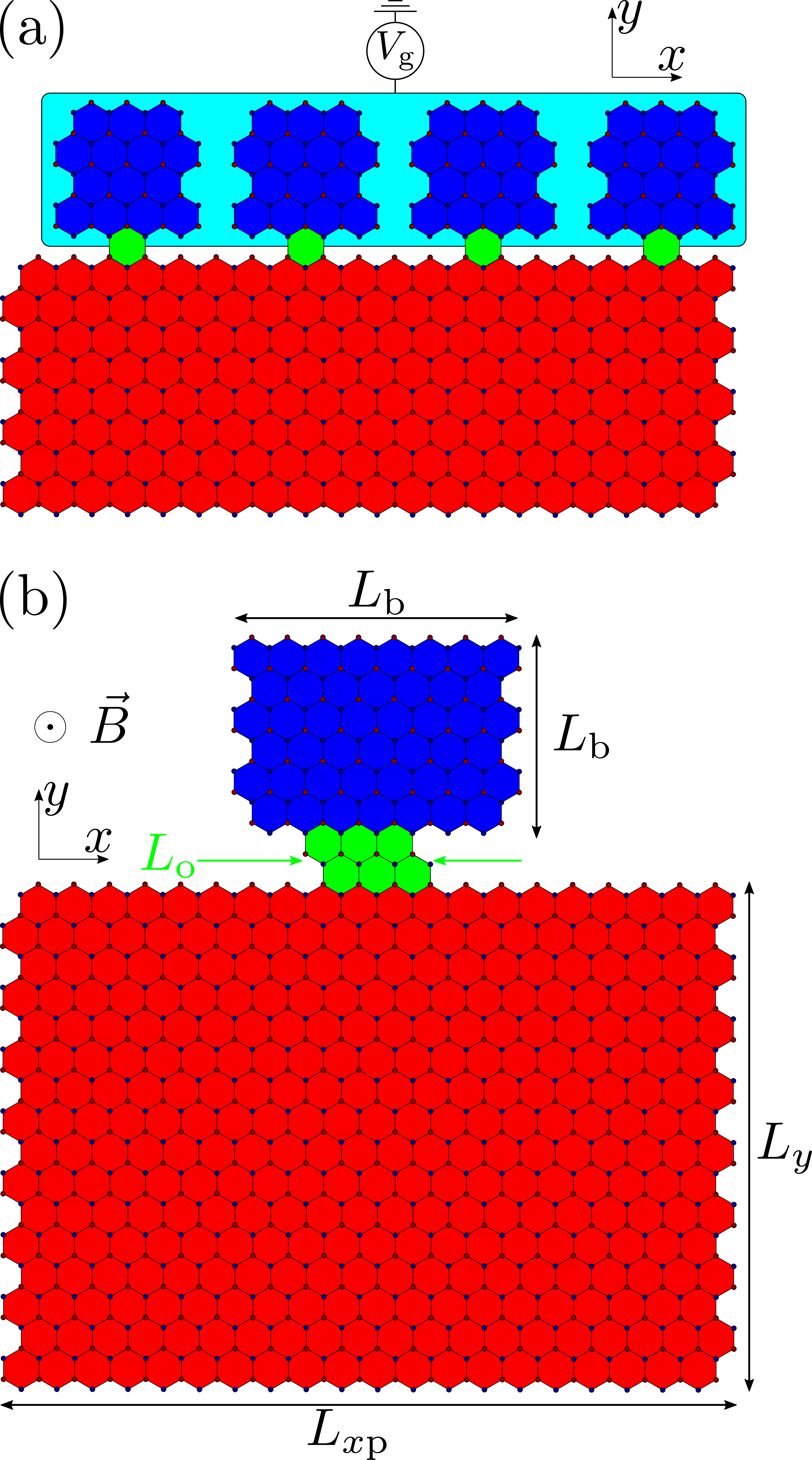}
	\caption{Panel (a): Tailored strip of graphene proposed for tunable Fermi velocities. 
	A magnetic field perpendicular to the sheet of graphene induces the IQHE. The sample has 
	periodically aligned bays illustrated here by the zoom with $4$ bays. 
	The gate voltage $V_\text{g}$ is applied only to the bays highlighted in blue. The red area corresponds to 
	the bulk of the strip and the green area to the opening between the bulk and a bay. Panel (b): Schematic sketch of a single unit cell with the notation for the
	linear dimensions of the bays and the strip. 
	The total number $N_x$ of unit cells  determines the total length of the sample 
	$L_x = N_x L_{x\mathrm{p}}$.}
	\label{fig:bay}
\end{figure}

The key challenge is to tailor the edges such that periodically arranged bays are weakly coupled 
to the chiral edge states, see Fig.\ \ref{fig:bay}(a). Due to the weak coupling controlled by the
width of the opening of the bays the local modes in the bays hybridize with the dispersive edge
modes. This hybridization leads to coupled modes with very little dispersion, hence very low
Fermi velocity. By applying gate voltages to adjust the chemical potential or the local potential of the bays the Fermi velocity is tuned: choosing the external parameters such that the edge mode is in resonance 
with the local bay modes reduces the Fermi velocity drastically. 

The paper is set up as follows. First, we introduce 
the model describing the edge states and the LL of graphene. 
Next, we discuss the  geometry of the considered samples. 
The main part shows the resulting dispersions and the tunability of the Fermi velocity 
in the edge states controlled by gate voltages as external control parameters. Finally, we conclude by summarizing and discussing possible applications.

\section{Model and Method}

The electronic properties of graphene in the vicinity of the Fermi energy 
are well reproduced by a fermionic tight-binding model. 
Due to the negligible contribution of the spin degree of freedom as well as of interactions 
we consider spinless fermions in the Hamiltonian 
\begin{equation}
	\mathcal{H} = t \sum_{\left\langle i, j \right\rangle} c_i^\dagger c_j^{\phantom{\dagger}} 
 - V_{\mathrm{g}} \sum_{i \in \text{bay}} 
	c_i^\dagger c_i^{\phantom{\dagger}}.
	\label{eq:hamiltonian}
\end{equation}
We focus on zero temperature so that the chemical potential $\mu$ is identical to the
Fermi energy $E_{\mathrm{F}}$. All states up to $\mu$ are occupied while the states above
$\mu$ are empty. The relevant Fermi velocity is the derivative of the dispersion at
the Fermi energy. A pair of nearest neighbors is denoted by $\left\langle i, j \right\rangle$. 
The tight-binding parameters are the nearest-neighbor hopping $t = \SI{2.8}{eV}$ and 
the lattice constant $a = \SI{0.142}{nm}$ \cite{liu15}.

In order to obtain the quantum Hall state in graphene we apply an external perpendicular 
magnetic field $\vec{B}$. This leads to the formation of LLs \cite{zheng02c, mcclu56} with 
energies at
\begin{equation}
	E_n = \mathrm{sgn}(n) \sqrt{2 e \hbar v_{\mathrm{F}}^2 B|n|} = 
	\mathrm{sgn}(n) \sqrt{\frac{9 t^2 a^2}{2 l_B^2} |n|}.
\end{equation}
The second equation stems from $\hbar v_{\mathrm{F}} = 3/2 t a$ and the definition of the magnetic length 
$l_B = \sqrt{h/(e B)} \approx 25.65564/\sqrt{B} \ \si{nm}$ where the of the magnetic field $B$ is inserted in units of Tesla. The non-linear spacing between 
the LL results from the relativistic behavior of the electrons near the Dirac points. 
The magnetic length plays the same role as in the IQHE in the 2DEG \cite{malki17c, delpl10}. 
It sets the scale for the diameter of the circular motion of the electrons due to the
Lorentz force. In order that the decoration of the edges by bays has an appreciable effect
the geometric dimensions of the bays must be of the order of this magnetic length.

The magnetic field $B$ is included in the tight-binding model by the Peierls substitution 
attributing  Aharanov--Bohm phases \cite{aharo59} to the hopping processes
\begin{equation}
t \rightarrow t \, \exp \left( \mathrm{i} e/\hbar \int_{r_1}^{r_2} \vec{A} \mathrm{d} \vec{r} \right),
\end{equation}
where the start and the end site of the hopping process is denoted by $r_1$ and $r_2$, respectively.
The evolution of the Dirac cones to rather flat LLs has been exhaustively studied 
and discussed by Delplace and Montambaux \cite{delpl10}. In order to keep 
translational invariance in the $x$-direction, see Fig.\ \ref{fig:bay}(a), we 
employ the Landau gauge $\vec{A}  = B(-y, 0, 0)$.
Thus, the momentum $k_x$ is a good quantum number in the calculations.

\begin{figure*}
	\centering
		\includegraphics[width=1\textwidth]{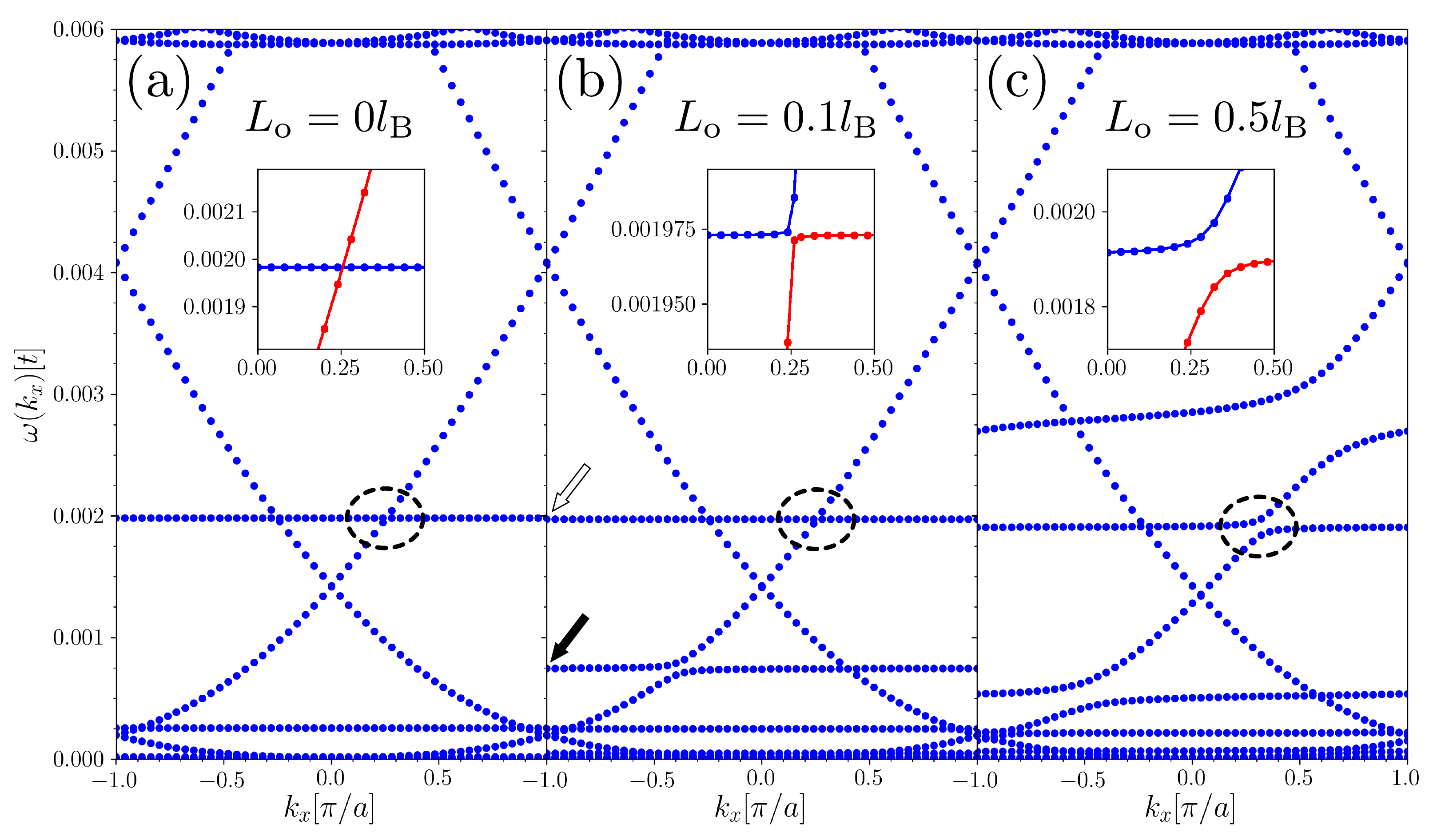}
	\caption{Dispersions of the lowest positive eigen energies in graphene in the IQHE at 
	$B = \SI{0.25}{T}, L_{x \mathrm{p}} = 3 l_B \approx \SI{153}{nm}$, 
	$L_y = 6 l_B \approx \SI{306}{nm}$, $L_\mathrm{b} = 2 l_B \approx \SI{102}{nm}$, and 
	$L_\mathrm{o} = \{ 0, 0.1, 0.5 \} l_B \approx \{0, 5, 26 \} \si{nm}$. 
	Panels (a-c) show the cases of uncoupled, weakly, and moderately coupled bays, respectively. 
Crossings evolve into avoided crossings due to the hybridization of both modes. The dashed ellipses mark a location where this happens.	
	The insets zoom into 
	the evolution of the avoided crossings at $k_x=\pi/(4a)$, where the two dispersion branches are highlighted in blue and red.}
	\label{fig:dispersion}
\end{figure*}

The numerical analysis is facilitated by small system size, i.e., the number of sites in the
extended unit cell, see Fig.\ \ref{fig:bay}(b), should be rather small so that
the dimension of the resulting eigen value problem remains tractable.
For the experimental realization, however, it is advantageous to consider rather large 
extended unit cells. In the following, we briefly discuss these constraints, the employed numerical
methods, and justify our choice of parameters.

The tight-binding Hamiltonian \eqref{eq:hamiltonian} comprises only on-site and nearest-neighbor 
hoppings. Hence, its very large matrix representing the Hamiltonian is mostly populated by zero 
entries and may therefore be encoded as sparse matrix. The signal transmission is 
primarily determined by the properties at the Fermi energy $E_\mathrm{F}$. 
Undoped graphene is a semi-metal with 
$E_{\mathrm{F}} = 0$. The states at higher energies hardly influence the low-energy dynamics.
So we focus on the low LLs up to the third one, i.e., with $|n| < 4$.
The eigen value solver FEAST \cite{poliz09} is used to constrain the considered
interval of the energy spectrum. This routine has proved \cite{malki17b} to implement a reliable 
high-performance algorithm to efficiently diagonalize large sparse matrices. It is based on the 
quantum mechanical density matrix representation and applies counter integration techniques 
in order to solve the eigen value problem in a fixed interval of the total spectrum.
Based on  the calculated eigen energies the Fermi velocity as derivative of the dispersion
at the Fermi energy is straightforwardly approximated by the ratio of finite differences.
We assume periodic boundary conditions in $x$-direction with $50$ unit cells. This implies
a sufficiently fine discretization of the Brillouin zone to display energy crossings and
avoided energy crossings.

The experimental constraints consist in the limitations in accurately tailoring the strips of graphene
with the desired structure at the edges, i.e., with the periodic structure of bays, while
maintaining a high mobility to realize the quantum Hall state. The bay pattern can be customized by 
electron beam lithography \cite{chen07b, hill06} or by anisotropic etching techniques \cite{campo09}. It appears that creating bays of the size of $\SI{100}{nm}$ can be done without major problems. 
In our calculations we assume quadratic bays for simplicity. The precise shape of the bays does
not matter for the qualitative results although it will have an influence on the
quantitative details. The size of the bays and especially the length of the bay opening 
$L_{\mathrm{o}}$ to the bulk of the strip, see Fig.\ \ref{fig:bay}(b),
 are crucial for dispersions of the modified edge states. To keep the example neat we 
aim at a small number of low-lying levels in the bays so that the relevant number of states which
may hybridize stays easily tractable. This implies $L_\text{b}\approx l_B$ and favors small
magnetic fields ($\SI{100}{mT}$ with $l_B \approx \SI{81}{nm}$).
These considerations define the framework for the numerical results in the next section.

\section{Results}

\subsection{Dispersions, hybridized edge modes, and localization}

In Fig.\ \ref{fig:dispersion}, three representative cases are depicted: uncoupled ($L_\text{o}=0l_B$), weakly 
($L_\text{o}=0.1l_B$), and moderately ($L_\text{o}=0.5l_B$) coupled bays. 
The magnetic field is set to $B = \SI{0.25}{T}$ which corresponds to a magnetic length 
$l_B \approx \SI{51}{nm}$.  The definitions of the various lengths are displayed in Fig.\ \ref{fig:bay}(b).
They are given by: 
$L_{x \mathrm{p}} = 3 l_B \approx \SI{153}{nm}$, $L_y = 6 l_B \approx \SI{306}{nm}$, 
$L_\mathrm{b} = 2 l_B \approx \SI{102}{nm}$ and $L_\mathrm{o} = \{ 0, 0.1, 0.5 \} l_B \approx \{0, 5, 26 \} 
\si{nm}$. The width of the strip $L_y$ is chosen large enough so that the two counter-propagating edge states
at the opposite edges do not overlap. This implies that both edges can be modified independent of each other. 
For clarity, we exploit this simplifying fact and modify only the upper edge while the lower edge remains a bare 
zigzag edge. The chosen bay size $L_\mathrm{b}$ should be experimentally realizable. 
The condition $L_{x \mathrm{p}} > L_\mathrm{b}$ ensures that the bays are separated from one another. 
Decreasing the distance between the bays by changing $L_{x\mathrm{p}}$ increases the impact on the dispersion of edge states due to the changed fraction of the  decorated boundary to the undecorated boundary. The three values for the opening $L_\mathrm{o}$ lead to three different degrees of hybridization.
As a rule of thumb a wider opening corresponds to a larger hybridization. 
The effect is discernible in the dispersions in Fig.\ \ref{fig:dispersion}. For the sake of clarity, 
we display the dispersions up to the first LL so that only two counter-propagating edge states need to be 
taken into account. 

Panel (a) in Fig.\ \ref{fig:dispersion} shows the uncoupled case where the usual LLs and their related  
edge states are distinct from the local states. The LLs are essentially flat and turn upwards where their
wave functions approach the edges of the strip. In contrast, the eigen states of the modes in the bays
are completely local, hence completely flat as function of the wave vector $k_x$. Thus, edge states and local modes show crossings, see inset of Fig.\ \ref{fig:dispersion}(a).
Due to the extended unit cell comprising one bay, the edge states are backfolded into the reduced 
Brillouin zone scheme \cite{malki17b}. Since the dispersion is  symmetric with respect to 
the $k_x$-axis, we only show the positive eigen energies. 
When the coupling of the bays to the bulk of the strip is switched on, i.e., $L_\mathrm{o} \neq 0$, see panels 
(b) and (c) in Fig.\ \ref{fig:dispersion}, the edge states mix with the local states. We observe that the crossings turn into avoided 
crossings with the bay modes due to level repulsion, see for instance the encircled regions marked by dashed ellipses. Increasing the 
opening further  and further results in a stronger and stronger  level repulsion so that the former 
local modes become more and more dispersive. 

\begin{figure}
	\centering
		\includegraphics[width=\columnwidth]{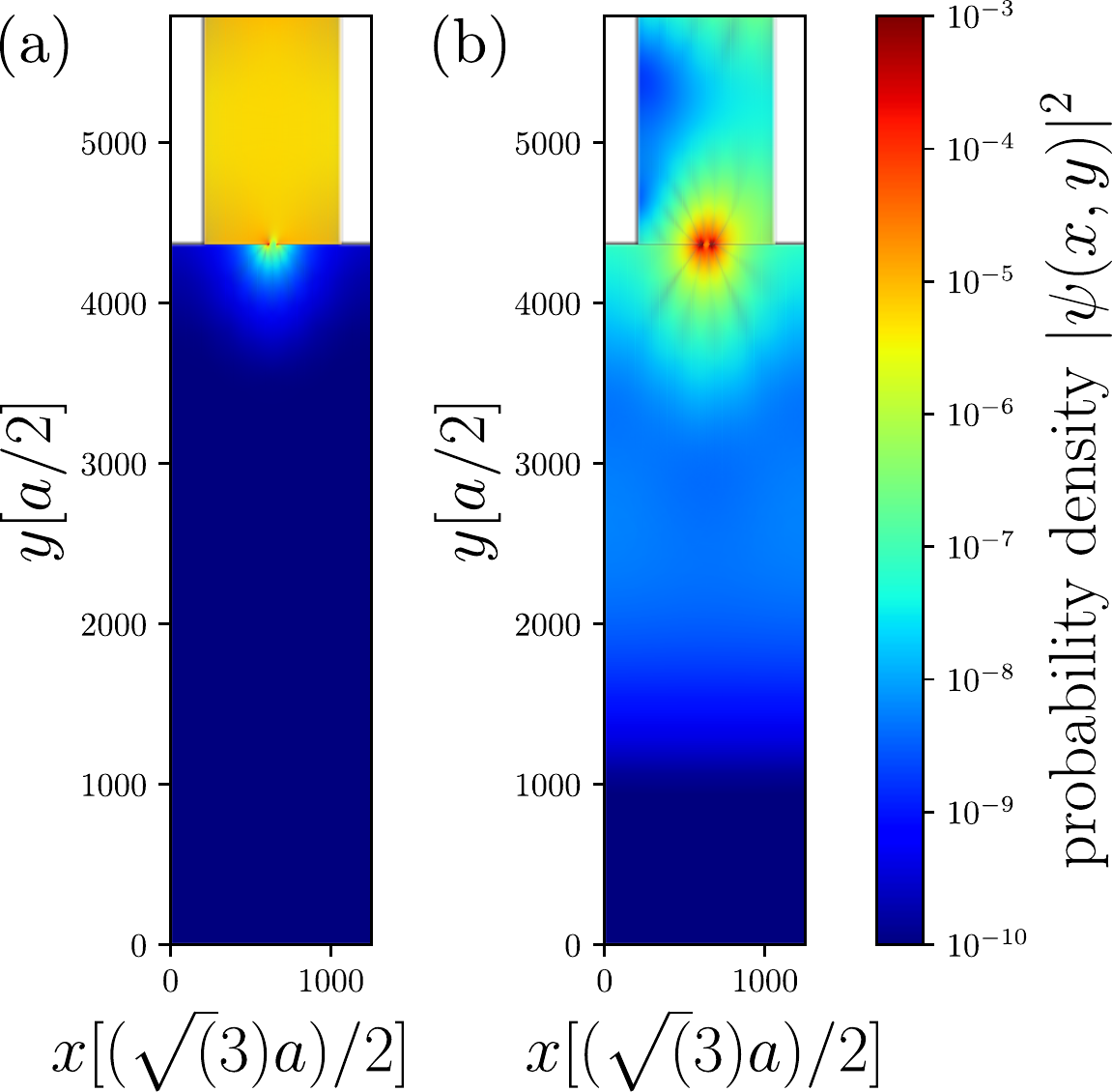}
	\caption{
	Probability densities $|\psi_n(x,y)|^2$ of two almost local eigen states at $k_x = 0$.
	Clearly, these modes are localized in two different regions. Panel (a) shows the local mode
	stemming from the bay whereas panel (b) shows the local mode localized in the opening of the bay. 
	The corresponding eigen states are indicated by open (mode from bay) and filled (mode in opening) arrows 
	in panel (b) of \mbox{Fig.\ \ref{fig:dispersion}.} The parameters of field and geometry are 
	$B = \SI{0.25}{T}, L_{x \mathrm{p}} = 3 l_B \approx \SI{153}{nm}$, $L_y = 6 l_B \approx \SI{306}{nm}$, 
	$L_\mathrm{b} = 2 l_B \approx \SI{102}{nm}$, and $L_\mathrm{o} = 0.1 l_B \approx \SI{5}{nm}$.} 
	\label{fig:psi}
\end{figure}

Much to our surprise, in addition to the local states of the bays other almost local states appear
upon opening the bays. Such local states have not been observed in the IQHE of 
the non-relativistic 2DEG \cite{malki17b}. 
Investigating the probability density of the eigen states at $k_x = 0$ reveals the location
 of the unexpected  modes. In \mbox{Fig.\ \ref{fig:psi}} we display the probability densities of the 
two states with the energies highlighted by arrows in panel (b) of \mbox{Fig.\ \ref{fig:dispersion}.} 
Figure \ref{fig:psi}(a) clearly shows the density of the local state from the bays marked in 
Fig.\ \ref{fig:dispersion}(b) by the open arrow: it is almost 
entirely localized within the bay and leaks only weakly into the bulk of the strip.
In contrast, \mbox{Fig.\ \ref{fig:psi}(b)} clearly shows a strong localization in the opening. 
This is the density of the additional local state marked in 
\mbox{Fig.\ \ref{fig:dispersion}(b)} by the filled arrow.
Obviously, the opening gives rise to additional localization. Hence, the added sites in the opening, see green area of Fig.\ \ref{fig:bay}(b), contribute to the spectra similar to the effect of the bay sites.
Independent of the origin of the almost local states, both hybridize with the edge modes at the same energy.
The additional state hybridizes more strongly as expected since it is localized in the opening very close
to the bulk of the strip while the mode from within the bay leaks only weakly into the bulk.
The hybridization of both local modes leads to a reduced Fermi velocity. 

% The hybridization between local LL from the bulk and the edge states are also present similar to the 
% case of the integer QHE for a free $2$D electron gas \cite{malki17b}. This effect however is of less 
% importance, since it is a second order effect and close to the bulk states which make it hard to 
% use for applications.

\subsection{Fermi velocities for signal transmission}

In order to tune the Fermi velocity of the edge state at the upper edge of the strip of graphene we have to
change the derivative of its dispersion at the Fermi level. To do so two ways suggest themselves in particular.
The most transparent one is to change the Fermi energy, i.e., the chemical potential, such that the 
Fermi level lies in a rather flat region of the edge state dispersion. This can be achieved by a gate close
to the total strip of graphene \cite{zhang05}. Alternatively, one may 
conceive a control of the energy level of the bays alone. This can be achieved by a voltage applied to an appropriate gate close to the edge of the strip, see Fig.\ \ref{fig:bay}(a). Undoubtedly, other ways of tuning 
can be devised as well. Below we present results for both approaches to show that tuning of the Fermi
velocity is possible.

\begin{figure*}
	\centering
		\includegraphics[width=1\textwidth]{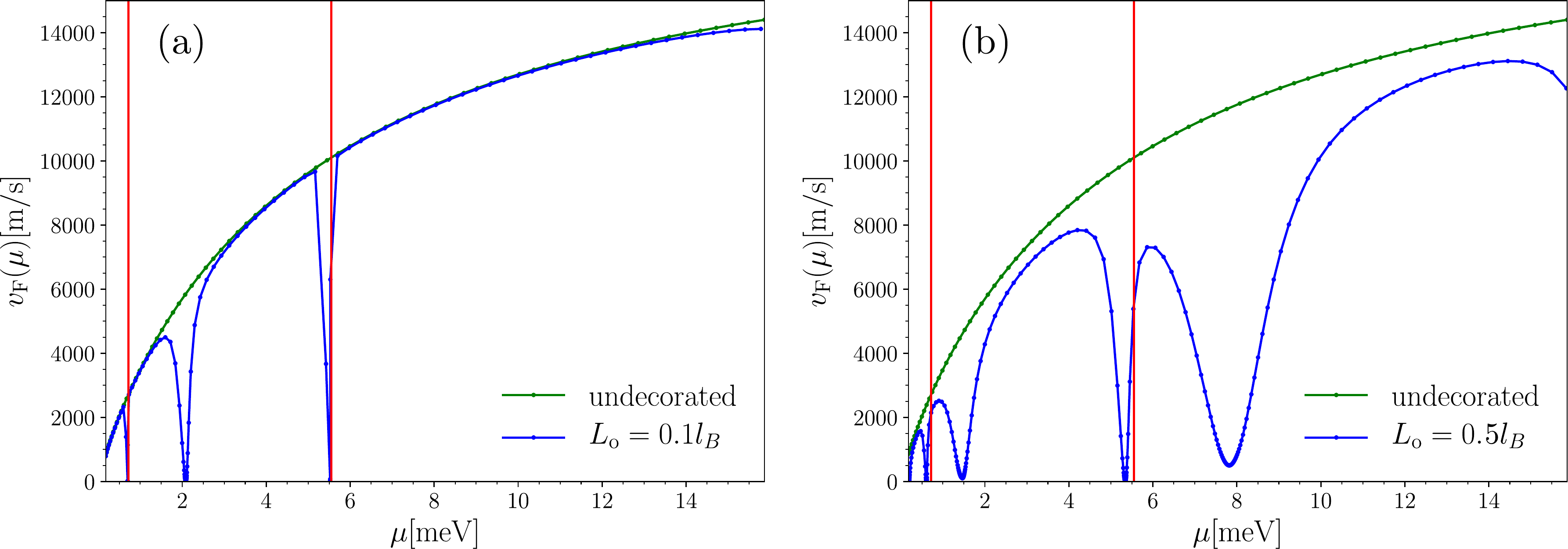}
	\caption{Fermi velocity $v_\mathrm{F}$ of the right-moving edge states localized at the upper edge
		as function of the chemical potential $\mu$ equivalent to the Fermi energy. 
		Panels (a) and (b) show the weakly and moderately coupled 
		cases $L_\mathrm{o} = \{0.1, 0.5 \} l_B \approx \{5, 26 \} \si{nm}$. 
		The other parameters are $B = \SI{0.25}{T}, L_{x \mathrm{p}} = 3 l_B \approx \SI{153}{nm}$, 
		$L_y = 6 l_B \approx \SI{306}{nm}$, $L_\mathrm{b} = 2 l_B \approx \SI{102}{nm}$.}
	\label{fig:vf_chem}
\end{figure*}

Figure \ref{fig:vf_chem} depicts the Fermi velocity as function of the chemical potential. 
Panel (a) and (b) correspond to the case of weakly and moderately coupled bays, respectively, of which
the dispersions are displayed in Fig.\ \ref{fig:dispersion}. The Fermi velocity in the weakly coupled case
shows three deep dips. The two extremely narrow and steep dips can be attributed to the hybridization with
two local levels from the isolated bays. The energetic position of these local levels is indicated by
vertical red lines in the two panels. The closeness of the steep dips to these lines underlines
their physical origin. The slight shifts of the dips relative to the red lines result from 
the energy shift due to the hybridization. This is supported by the fact that the shift is larger
for panel (b) which refers to bays with a wider opening and hence stronger hybridization.
The broader dips (one in panel (a) and two in panel (b)) are related to the additional local states 
localized in the openings, see Fig.\ \ref{fig:psi}(b). The fact that these dips are broader is explained 
by the vicinity of these states to the bulk of the strips implying a stronger hybridization
with the edge mode than for the local states from within the bays.

By tuning $\mu$ into these dips the Fermi velocity can by reduced by orders of magnitude.
In particular for narrow bay openings a strong reduction can be achieved. 
For instance, $v_\mathrm{F}$ in panel (a) of Fig.\ \ref{fig:vf_chem} at $\mu \approx \SI{2}{meV}$ is reduced 
by a factor of $ \approx 65 000$ from its value without tailored edges, i.e., without bays, see green curve in
Fig.\ \ref{fig:vf_chem}. Increasing the bay opening leads to a broadening and a shift of the dips, 
see Fig.\ \ref{fig:vf_chem}(b). The shift of the local modes results from the hybridization pushing the local modes downwards in energy due to level repulsion. Furthermore, a wider opening possesses more low-lying 
energy modes so that the number of dips is increased. 
If the chemical potential is out-of-resonance with local modes, 
the reduction of the Fermi velocity is insignificant. This is particularly true for narrow opening where
there is no significant difference between the green and the blue line out-of-resonance.

\begin{figure}
	\centering
		\includegraphics[width=1\columnwidth]{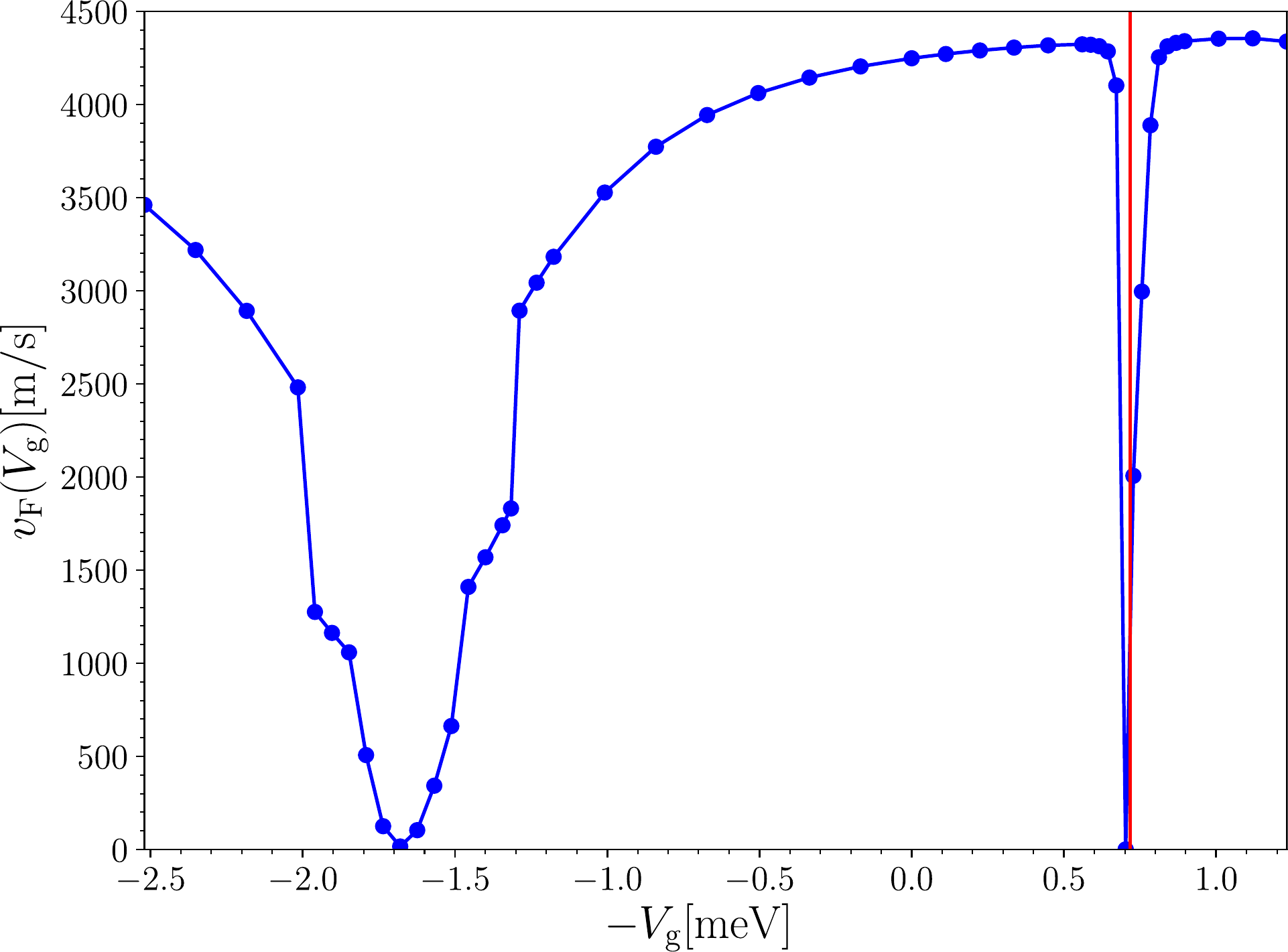}
	\caption{Fermi velocity $v_\mathrm{F}$ of the right-moving edge states as function of the gate voltage
	$V_\mathrm{g}$ changing only the energy offset of the bays, see Fig.\ \ref{fig:bay}(a). The parameters
	are $B = \SI{0.25}{T}, L_{x \mathrm{p}} = 3 l_B \approx \SI{153}{nm}$, $L_y = 6 l_B \approx \SI{306}{nm}$, 
	$L_\mathrm{b} = 2 l_B \approx \SI{102}{nm}$, and $L_\mathrm{o} = 0.1 l_B \approx \SI{5}{nm}$. The less smooth 
	dip at about $V_\mathrm{g}=-1.7$meV results from the discretization of the Brillouin zone with 50 points. 
	A finer	mesh would lead to a smoother curve, but is computationally demanding.}
	\label{fig:vf_bay}
\end{figure}

Tuning of the bay potential $V_\mathrm{g}$ leads to similar results even though changing this gate voltage 
does not simply shift the local modes relative to the bulk modes.
Still, significant changes of the Fermi velocity can be realized as illustrated in Fig.\
\ref{fig:vf_bay} for which we assume a generic Fermi energy $E_\mathrm{F} = 0.0005 t = \SI{1.4}{meV}$ 
in the regime of doped graphene. The finite value of the Fermi energy is necessary to be in the dispersive 
regime of the edge states. In order to realize a substantial reduction of the Fermi velocity, the local 
states must be brought into resonance with the edge states. This tuning of the bay potential results in
similar dips of the Fermi velocity as the tuning of the chemical potential $\mu$. 
Negative values of $V_\mathrm{g}$ lift the bay spectrum up in energy so that they come into resonance
with the edge state resulting in a steep dip.

In contrast, a positive gate voltage $V_\mathrm{g}$ decreases the energy of the additional local model in 
the opening. The ensuing resonance leads to a broader dip due to the larger hybridization of the opening 
mode to the edge mode. Note that larger absolute values $ V_\mathrm{g}$ are needed to reach the resonance
compared to what can be estimated from the dispersion. This fact can be explained by the level repulsion 
from the the bulk states which are mixed in. 
Thus, both dips are shifted relative to the energy differences in the uncoupled dispersion, 
see panel (a) in \mbox{Fig.\ \ref{fig:dispersion}.} 
We conclude that  tuning the bay potential is also a suitable knob to control the Fermi velocity. 
In summary, tuning both gate voltages leads to resonances with local modes so that 
substantial reductions of the velocity of signal transmission are achieved.

\section{Summary and discussion}

The above results clearly show that substantial tuning of the Fermi velocity in graphene with tailored
edges is possible. This can be used to construct tunable delay lines for charge signals: 
by controlling the external parameters such as gate voltages the temporal delay can be tuned at will
by choosing an appropriate velocity of signal transmission through the sample. 
This may help to control and to switch properties of nanoscale devices including devices for
quantum information processes. The delayed signal itself does not need to have quantum character,
but the tuned delay can help to deliver control precisely at required time instants.

Another promising idea in the same spirit is to construct interferometers in which two signals
are superposed which have propagated along different pathways. One of these pathways contains
the graphene sheet with tunable signal velocity. By adjusting the Fermi velocity the delay
in this path can be altered such that destructive or constructive interference takes place.
If the signal along the other pathway is propagating through a sample of an unknown 
compound or an unknown device its transmission properties can be investigated in this way.
These two suggestions are meant to exemplify promising applications of tunable signal velocities.
We emphasize that the tuning can be done very fast on the time scales on which the gate
voltages can be changed. 

To obtain an idea of the order of magnitude of the delays we assume the conditions used for 
Fig.\ \ref{fig:vf_chem}(b), but with only $N_x = 20$. For simplicity, rounded values are used. 
The sample length is $L_x \approx \SI{3}{\mu m}$ and we consider the broader dip because its
less susceptible to imperfections. For $\mu = \SI{8}{\meV}$ the undecorated strip of graphene
displays a velocity $v_\mathrm{F} = \SI{12 000}{m/s}$. It will be reduced by a factor of $20$ 
down to $v_\mathrm{F} = \SI{600}{m/s}$ in the minimum of the dip. The time required by a signal 
to propagate along the edge is delayed from $\SI{0.25}{ns}$ to $\SI{5}{ns}$. 
This change of transmission time is readily detectable.

We are aware that the calculations presented in this article assume idealized conditions, for instance zero
temperature and small samples without imperfections or disorder. So further research is called for to
address these points as we did already for the
non-relativistic IQHE \cite{malki17b}. Nevertheless, the origin of the predicted effect is clearly elucidated and
does not rely on subtleties of the model, for instance the shape of the bays does not
matter as long as the bays host local modes. With state-of-the-art techniques the IQHE can be detected in 
graphene samples which are not smaller than $3 \times \SI{3}{\micro\meter^2}$. For smaller samples, 
imperfections and disorder spoil the Hall states. As long as the IQHE can be observed and the properties of
at least one of the edges can be externally controlled tuning of the signal velocity will be within
reach.

The take-home message is that graphene represents a promising material to realize the IQHE with
externally tunable dispersions of the edge states on the basis of today's technology. 
Tailoring of the edges is an exacting prerequisite. We hope that our results trigger further studies,
in particular experimental ones, paving the way towards tunable signal velocities. As an outlook
we point out that further work on the influence of imperfections, disorder, and the presence of
the spin degree of freedom are in order. The latter opens up particular applications in spintronics.

\acknowledgments

This work was supported by the Deutsche Forschungsgemeinschaft and the Russian Foundation of Basic Research 
in TRR 160. MM gratefully acknowledges financial support by the Studienstiftung des deutschen Volkes. 
GSU thanks Alex R.\ Hamilton and Oleg P.\ Sushkov for useful discussions
and the Heinrich-Hertz Stiftung for enabling these discussions by financial support.

%\bibliographystyle{apsrev4-1}
%\bibliography{../../bibinput/liter10}
%\bibliography{liter10}

\begin{thebibliography}{32}%
\makeatletter
\providecommand \@ifxundefined [1]{%
 \@ifx{#1\undefined}
}%
\providecommand \@ifnum [1]{%
 \ifnum #1\expandafter \@firstoftwo
 \else \expandafter \@secondoftwo
 \fi
}%
\providecommand \@ifx [1]{%
 \ifx #1\expandafter \@firstoftwo
 \else \expandafter \@secondoftwo
 \fi
}%
\providecommand \natexlab [1]{#1}%
\providecommand \enquote  [1]{``#1''}%
\providecommand \bibnamefont  [1]{#1}%
\providecommand \bibfnamefont [1]{#1}%
\providecommand \citenamefont [1]{#1}%
\providecommand \href@noop [0]{\@secondoftwo}%
\providecommand \href [0]{\begingroup \@sanitize@url \@href}%
\providecommand \@href[1]{\@@startlink{#1}\@@href}%
\providecommand \@@href[1]{\endgroup#1\@@endlink}%
\providecommand \@sanitize@url [0]{\catcode `\\12\catcode `\$12\catcode
  `\&12\catcode `\#12\catcode `\^12\catcode `\_12\catcode `\%12\relax}%
\providecommand \@@startlink[1]{}%
\providecommand \@@endlink[0]{}%
\providecommand \url  [0]{\begingroup\@sanitize@url \@url }%
\providecommand \@url [1]{\endgroup\@href {#1}{\urlprefix }}%
\providecommand \urlprefix  [0]{URL }%
\providecommand \Eprint [0]{\href }%
\providecommand \doibase [0]{http://dx.doi.org/}%
\providecommand \selectlanguage [0]{\@gobble}%
\providecommand \bibinfo  [0]{\@secondoftwo}%
\providecommand \bibfield  [0]{\@secondoftwo}%
\providecommand \translation [1]{[#1]}%
\providecommand \BibitemOpen [0]{}%
\providecommand \bibitemStop [0]{}%
\providecommand \bibitemNoStop [0]{.\EOS\space}%
\providecommand \EOS [0]{\spacefactor3000\relax}%
\providecommand \BibitemShut  [1]{\csname bibitem#1\endcsname}%
\let\auto@bib@innerbib\@empty
%</preamble>
\bibitem [{\citenamefont {Thouless}\ \emph {et~al.}(1982)\citenamefont
  {Thouless}, \citenamefont {Kohmoto}, \citenamefont {Nightingale},\ and\
  \citenamefont {den Nijs}}]{thoul82}%
  \BibitemOpen
  \bibfield  {author} {\bibinfo {author} {\bibfnamefont {D.~J.}\ \bibnamefont
  {Thouless}}, \bibinfo {author} {\bibfnamefont {M.}~\bibnamefont {Kohmoto}},
  \bibinfo {author} {\bibfnamefont {M.~P.}\ \bibnamefont {Nightingale}}, \ and\
  \bibinfo {author} {\bibfnamefont {M.}~\bibnamefont {den Nijs}},\ }\href
  {\doibase 10.1103/PhysRevLett.49.405} {\bibfield  {journal} {\bibinfo
  {journal} {Phys. Rev. Lett.}\ }\textbf {\bibinfo {volume} {49}},\ \bibinfo
  {pages} {405} (\bibinfo {year} {1982})}\BibitemShut {NoStop}%
\bibitem [{\citenamefont {Avron}\ \emph {et~al.}(1983)\citenamefont {Avron},
  \citenamefont {Seiler},\ and\ \citenamefont {Simon}}]{avron83}%
  \BibitemOpen
  \bibfield  {author} {\bibinfo {author} {\bibfnamefont {J.~E.}\ \bibnamefont
  {Avron}}, \bibinfo {author} {\bibfnamefont {R.}~\bibnamefont {Seiler}}, \
  and\ \bibinfo {author} {\bibfnamefont {B.}~\bibnamefont {Simon}},\ }\href
  {\doibase 10.1103/PhysRevLett.51.51.} {\bibfield  {journal} {\bibinfo
  {journal} {Phys. Rev. Lett.}\ }\textbf {\bibinfo {volume} {51}},\ \bibinfo
  {pages} {51} (\bibinfo {year} {1983})}\BibitemShut {NoStop}%
\bibitem [{\citenamefont {Niu}\ \emph {et~al.}(1985)\citenamefont {Niu},
  \citenamefont {Thouless},\ and\ \citenamefont {Wu}}]{niu85}%
  \BibitemOpen
  \bibfield  {author} {\bibinfo {author} {\bibfnamefont {Q.}~\bibnamefont
  {Niu}}, \bibinfo {author} {\bibfnamefont {D.~J.}\ \bibnamefont {Thouless}}, \
  and\ \bibinfo {author} {\bibfnamefont {Y.-S.}\ \bibnamefont {Wu}},\ }\href
  {\doibase 10.1103/PhysRevB.31.3372} {\bibfield  {journal} {\bibinfo
  {journal} {Phys. Rev. B}\ }\textbf {\bibinfo {volume} {31}},\ \bibinfo
  {pages} {3372} (\bibinfo {year} {1985})}\BibitemShut {NoStop}%
\bibitem [{\citenamefont {Kohmoto}(1985)}]{kohmo85}%
  \BibitemOpen
  \bibfield  {author} {\bibinfo {author} {\bibfnamefont {M.}~\bibnamefont
  {Kohmoto}},\ }\href@noop {} {\bibfield  {journal} {\bibinfo  {journal} {Ann.
  of Phys.}\ }\textbf {\bibinfo {volume} {160}},\ \bibinfo {pages} {343}
  (\bibinfo {year} {1985})}\BibitemShut {NoStop}%
\bibitem [{\citenamefont {Weis}\ and\ \citenamefont {von
  Klitzing}(2011)}]{weis11}%
  \BibitemOpen
  \bibfield  {author} {\bibinfo {author} {\bibfnamefont {J.}~\bibnamefont
  {Weis}}\ and\ \bibinfo {author} {\bibfnamefont {K.}~\bibnamefont {von
  Klitzing}},\ }\href {\doibase 10.1098/rsta.2011.0198} {\bibfield  {journal}
  {\bibinfo  {journal} {Phil. Trans. R. Soc. A}\ }\textbf {\bibinfo {volume}
  {369}},\ \bibinfo {pages} {3954} (\bibinfo {year} {2011})}\BibitemShut
  {NoStop}%
\bibitem [{\citenamefont {Kadel}\ \emph {et~al.}(2011)\citenamefont {Kadel},
  \citenamefont {Kumari}, \citenamefont {Li}, \citenamefont {Huang},\ and\
  \citenamefont {Provencio}}]{kadel11}%
  \BibitemOpen
  \bibfield  {author} {\bibinfo {author} {\bibfnamefont {K.}~\bibnamefont
  {Kadel}}, \bibinfo {author} {\bibfnamefont {L.}~\bibnamefont {Kumari}},
  \bibinfo {author} {\bibfnamefont {W.}~\bibnamefont {Li}}, \bibinfo {author}
  {\bibfnamefont {J.~Y.}\ \bibnamefont {Huang}}, \ and\ \bibinfo {author}
  {\bibfnamefont {P.~P.}\ \bibnamefont {Provencio}},\ }\href {\doibase
  10.1007/s11671-010-9795-7} {\bibfield  {journal} {\bibinfo  {journal}
  {Nanoscale Res. Lett.}\ }\textbf {\bibinfo {volume} {6}},\ \bibinfo {pages}
  {57} (\bibinfo {year} {2011})}\BibitemShut {NoStop}%
\bibitem [{\citenamefont {Perry}(2011)}]{perry11}%
  \BibitemOpen
  \bibfield  {author} {\bibinfo {author} {\bibfnamefont {D.~L.}\ \bibnamefont
  {Perry}},\ }\href {\doibase 10.1201/b10908} {\emph {\bibinfo {title}
  {Handbook of Inorganic Compounds}}}\ (\bibinfo  {publisher} {CRC Press},\
  \bibinfo {address} {Boca Raton, FL, USA},\ \bibinfo {year}
  {2011})\BibitemShut {NoStop}%
\bibitem [{\citenamefont {Uhrig}(2016)}]{uhrig16}%
  \BibitemOpen
  \bibfield  {author} {\bibinfo {author} {\bibfnamefont {G.~S.}\ \bibnamefont
  {Uhrig}},\ }\href {\doibase 10.1103/PhysRevB.93.205438} {\bibfield  {journal}
  {\bibinfo  {journal} {Phys. Rev. B}\ }\textbf {\bibinfo {volume} {93}},\
  \bibinfo {pages} {205438} (\bibinfo {year} {2016})}\BibitemShut {NoStop}%
\bibitem [{\citenamefont {Malki}\ and\ \citenamefont
  {Uhrig}(2017{\natexlab{a}})}]{malki17b}%
  \BibitemOpen
  \bibfield  {author} {\bibinfo {author} {\bibfnamefont {M.}~\bibnamefont
  {Malki}}\ and\ \bibinfo {author} {\bibfnamefont {G.~S.}\ \bibnamefont
  {Uhrig}},\ }\href {\doibase 10.1103/PhysRevB.95.235118} {\bibfield  {journal}
  {\bibinfo  {journal} {Phys. Rev. B}\ }\textbf {\bibinfo {volume} {95}},\
  \bibinfo {pages} {235118} (\bibinfo {year} {2017}{\natexlab{a}})}\BibitemShut
  {NoStop}%
\bibitem [{\citenamefont {Malki}\ and\ \citenamefont
  {Uhrig}(2017{\natexlab{b}})}]{malki17c}%
  \BibitemOpen
  \bibfield  {author} {\bibinfo {author} {\bibfnamefont {M.}~\bibnamefont
  {Malki}}\ and\ \bibinfo {author} {\bibfnamefont {G.~S.}\ \bibnamefont
  {Uhrig}},\ }\href {\doibase 10.21468/SciPostPhys.3.4.032} {\bibfield
  {journal} {\bibinfo  {journal} {SciPost Phys.}\ }\textbf {\bibinfo {volume}
  {3}},\ \bibinfo {pages} {032} (\bibinfo {year}
  {2017}{\natexlab{b}})}\BibitemShut {NoStop}%
\bibitem [{\citenamefont {Zhang}\ \emph {et~al.}(2005)\citenamefont {Zhang},
  \citenamefont {Tan}, \citenamefont {Stormer},\ and\ \citenamefont
  {Kim}}]{zhang05}%
  \BibitemOpen
  \bibfield  {author} {\bibinfo {author} {\bibfnamefont {Y.}~\bibnamefont
  {Zhang}}, \bibinfo {author} {\bibfnamefont {Y.-W.}\ \bibnamefont {Tan}},
  \bibinfo {author} {\bibfnamefont {H.~L.}\ \bibnamefont {Stormer}}, \ and\
  \bibinfo {author} {\bibfnamefont {P.}~\bibnamefont {Kim}},\ }\href {\doibase
  10.1038/nature04235} {\bibfield  {journal} {\bibinfo  {journal} {Nature}\
  }\textbf {\bibinfo {volume} {438}},\ \bibinfo {pages} {201} (\bibinfo {year}
  {2005})}\BibitemShut {NoStop}%
\bibitem [{\citenamefont {Neto}\ \emph {et~al.}(2009)\citenamefont {Neto},
  \citenamefont {Guinea}, \citenamefont {Peres}, \citenamefont {Novoselov},\
  and\ \citenamefont {Geim}}]{neto09}%
  \BibitemOpen
  \bibfield  {author} {\bibinfo {author} {\bibfnamefont {A.~H.}\
  \bibnamefont {Castro~Neto}}, \bibinfo {author} {\bibfnamefont {F.}~\bibnamefont
  {Guinea}}, \bibinfo {author} {\bibfnamefont {N.~M.~R.}\ \bibnamefont
  {Peres}}, \bibinfo {author} {\bibfnamefont {K.~S.}\ \bibnamefont
  {Novoselov}}, \ and\ \bibinfo {author} {\bibfnamefont {A.~K.}\ \bibnamefont
  {Geim}},\ }\href {\doibase 10.1103/RevModPhys.81.109} {\bibfield  {journal}
  {\bibinfo  {journal} {Rev. Mod. Phys.}\ }\textbf {\bibinfo {volume} {81}},\
  \bibinfo {pages} {109} (\bibinfo {year} {2009})}\BibitemShut {NoStop}%
\bibitem [{\citenamefont {Novoselov}\ \emph {et~al.}(2005)\citenamefont
  {Novoselov}, \citenamefont {Geim}, \citenamefont {Morozov}, \citenamefont
  {Jiang}, \citenamefont {Katsnelson}, \citenamefont {Grigorieva},
  \citenamefont {Dubonos},\ and\ \citenamefont {Firsov}}]{novos05}%
  \BibitemOpen
  \bibfield  {author} {\bibinfo {author} {\bibfnamefont {K.~S.}\ \bibnamefont
  {Novoselov}}, \bibinfo {author} {\bibfnamefont {A.~K.}\ \bibnamefont {Geim}},
  \bibinfo {author} {\bibfnamefont {S.~V.}\ \bibnamefont {Morozov}}, \bibinfo
  {author} {\bibfnamefont {D.}~\bibnamefont {Jiang}}, \bibinfo {author}
  {\bibfnamefont {M.~I.}\ \bibnamefont {Katsnelson}}, \bibinfo {author}
  {\bibfnamefont {I.~V.}\ \bibnamefont {Grigorieva}}, \bibinfo {author}
  {\bibfnamefont {S.~V.}\ \bibnamefont {Dubonos}}, \ and\ \bibinfo {author}
  {\bibfnamefont {A.~A.}\ \bibnamefont {Firsov}},\ }\href {\doibase
  10.1038/nature04233} {\bibfield  {journal} {\bibinfo  {journal} {Nature}\
  }\textbf {\bibinfo {volume} {438}},\ \bibinfo {pages} {197} (\bibinfo {year}
  {2005})}\BibitemShut {NoStop}%
\bibitem [{\citenamefont {Geim}\ and\ \citenamefont
  {Novoselov}(2007)}]{geim07}%
  \BibitemOpen
  \bibfield  {author} {\bibinfo {author} {\bibfnamefont {A.~K.}\ \bibnamefont
  {Geim}}\ and\ \bibinfo {author} {\bibfnamefont {K.~S.}\ \bibnamefont
  {Novoselov}},\ }\href {\doibase 10.1038/nmat1849} {\bibfield  {journal}
  {\bibinfo  {journal} {Nat. Mater.}\ }\textbf {\bibinfo {volume} {6}},\
  \bibinfo {pages} {183} (\bibinfo {year} {2007})}\BibitemShut {NoStop}%
\bibitem [{\citenamefont {DiVincenzo}\ and\ \citenamefont
  {Mele}(1984)}]{divin84}%
  \BibitemOpen
  \bibfield  {author} {\bibinfo {author} {\bibfnamefont {D.~P.}\ \bibnamefont
  {DiVincenzo}}\ and\ \bibinfo {author} {\bibfnamefont {E.~J.}\ \bibnamefont
  {Mele}},\ }\href {\doibase 10.1103/PhysRevB.29.1685} {\bibfield  {journal}
  {\bibinfo  {journal} {Phys. Rev. B}\ }\textbf {\bibinfo {volume} {29}},\
  \bibinfo {pages} {1685} (\bibinfo {year} {1984})}\BibitemShut {NoStop}%
\bibitem [{\citenamefont {Semenoff}(1984)}]{semen84}%
  \BibitemOpen
  \bibfield  {author} {\bibinfo {author} {\bibfnamefont {G.~W.}\ \bibnamefont
  {Semenoff}},\ }\href {\doibase 10.1103/PhysRevLett.53.2449} {\bibfield
  {journal} {\bibinfo  {journal} {Phys. Rev. Lett.}\ }\textbf {\bibinfo
  {volume} {53}},\ \bibinfo {pages} {2449} (\bibinfo {year}
  {1984})}\BibitemShut {NoStop}%
\bibitem [{\citenamefont {Hwang}\ \emph {et~al.}(2012)\citenamefont {Hwang},
  \citenamefont {Siegel}, \citenamefont {Mo}, \citenamefont {Regan},
  \citenamefont {Ismach}, \citenamefont {Zhang}, \citenamefont {Zettl},\ and\
  \citenamefont {Lanzara}}]{hwang12b}%
  \BibitemOpen
  \bibfield  {author} {\bibinfo {author} {\bibfnamefont {C.}~\bibnamefont
  {Hwang}}, \bibinfo {author} {\bibfnamefont {D.~A.}\ \bibnamefont {Siegel}},
  \bibinfo {author} {\bibfnamefont {S.-K.}\ \bibnamefont {Mo}}, \bibinfo
  {author} {\bibfnamefont {W.}~\bibnamefont {Regan}}, \bibinfo {author}
  {\bibfnamefont {A.}~\bibnamefont {Ismach}}, \bibinfo {author} {\bibfnamefont
  {Y.}~\bibnamefont {Zhang}}, \bibinfo {author} {\bibfnamefont
  {A.}~\bibnamefont {Zettl}}, \ and\ \bibinfo {author} {\bibfnamefont
  {A.}~\bibnamefont {Lanzara}},\ }\href {\doibase 10.1038/srep00590} {\bibfield
   {journal} {\bibinfo  {journal} {Scientific reports}\ }\textbf {\bibinfo
  {volume} {2}},\ \bibinfo {pages} {590} (\bibinfo {year} {2012})}\BibitemShut
  {NoStop}%
\bibitem [{\citenamefont {Liu}\ \emph {et~al.}(2015)\citenamefont {Liu},
  \citenamefont {Rickhaus}, \citenamefont {Makk}, \citenamefont
  {T{\'o}v{\'a}ri}, \citenamefont {Maurand}, \citenamefont {Tkatschenko},
  \citenamefont {Weiss}, \citenamefont {Sch{\"o}nenberger},\ and\ \citenamefont
  {Richter}}]{liu15}%
  \BibitemOpen
  \bibfield  {author} {\bibinfo {author} {\bibfnamefont {M.-H.}\ \bibnamefont
  {Liu}}, \bibinfo {author} {\bibfnamefont {P.}~\bibnamefont {Rickhaus}},
  \bibinfo {author} {\bibfnamefont {P.}~\bibnamefont {Makk}}, \bibinfo {author}
  {\bibfnamefont {E.}~\bibnamefont {T{\'o}v{\'a}ri}}, \bibinfo {author}
  {\bibfnamefont {R.}~\bibnamefont {Maurand}}, \bibinfo {author} {\bibfnamefont
  {F.}~\bibnamefont {Tkatschenko}}, \bibinfo {author} {\bibfnamefont
  {M.}~\bibnamefont {Weiss}}, \bibinfo {author} {\bibfnamefont
  {C.}~\bibnamefont {Sch{\"o}nenberger}}, \ and\ \bibinfo {author}
  {\bibfnamefont {K.}~\bibnamefont {Richter}},\ }\href {\doibase
  10.1103/PhysRevLett.114.036601} {\bibfield  {journal} {\bibinfo  {journal}
  {Phys. Rev. Lett.}\ }\textbf {\bibinfo {volume} {114}},\ \bibinfo {pages}
  {036601} (\bibinfo {year} {2015})}\BibitemShut {NoStop}%
\bibitem [{\citenamefont {Gusynin}\ and\ \citenamefont
  {Sharapov}(2005)}]{gusyn05}%
  \BibitemOpen
  \bibfield  {author} {\bibinfo {author} {\bibfnamefont {V.~P.}\ \bibnamefont
  {Gusynin}}\ and\ \bibinfo {author} {\bibfnamefont {S.~G.}\ \bibnamefont
  {Sharapov}},\ }\href {\doibase 10.1103/PhysRevLett.95.146801} {\bibfield
  {journal} {\bibinfo  {journal} {Phys. Rev. Lett.}\ }\textbf {\bibinfo
  {volume} {95}},\ \bibinfo {pages} {146801} (\bibinfo {year}
  {2005})}\BibitemShut {NoStop}%
\bibitem [{\citenamefont {Bolotin}\ \emph {et~al.}(2008)\citenamefont
  {Bolotin}, \citenamefont {Sikes}, \citenamefont {Jiang}, \citenamefont
  {Klima}, \citenamefont {Fudenberg}, \citenamefont {Hone}, \citenamefont
  {Kim},\ and\ \citenamefont {Stormer}}]{bolot08}%
  \BibitemOpen
  \bibfield  {author} {\bibinfo {author} {\bibfnamefont {K.~I.}\ \bibnamefont
  {Bolotin}}, \bibinfo {author} {\bibfnamefont {K.~J.}\ \bibnamefont {Sikes}},
  \bibinfo {author} {\bibfnamefont {Z.}~\bibnamefont {Jiang}}, \bibinfo
  {author} {\bibfnamefont {M.}~\bibnamefont {Klima}}, \bibinfo {author}
  {\bibfnamefont {G.}~\bibnamefont {Fudenberg}}, \bibinfo {author}
  {\bibfnamefont {J.}~\bibnamefont {Hone}}, \bibinfo {author} {\bibfnamefont
  {P.}~\bibnamefont {Kim}}, \ and\ \bibinfo {author} {\bibfnamefont {H.~L.}\
  \bibnamefont {Stormer}},\ }\href {\doibase 10.1016/j.ssc.2008.02.024}
  {\bibfield  {journal} {\bibinfo  {journal} {Solid State Commun.}\ }\textbf
  {\bibinfo {volume} {146}},\ \bibinfo {pages} {351} (\bibinfo {year}
  {2008})}\BibitemShut {NoStop}%
\bibitem [{\citenamefont {Dean}\ \emph {et~al.}(2010)\citenamefont {Dean},
  \citenamefont {Young}, \citenamefont {Meric}, \citenamefont {Lee},
  \citenamefont {Wang}, \citenamefont {Sorgenfrei}, \citenamefont {Watanabe},
  \citenamefont {Taniguchi}, \citenamefont {Kim}, \citenamefont {Shepard},\
  and\ \citenamefont {Hone}}]{dean10}%
  \BibitemOpen
  \bibfield  {author} {\bibinfo {author} {\bibfnamefont {C.~R.}\ \bibnamefont
  {Dean}}, \bibinfo {author} {\bibfnamefont {A.~F.}\ \bibnamefont {Young}},
  \bibinfo {author} {\bibfnamefont {I.}~\bibnamefont {Meric}}, \bibinfo
  {author} {\bibfnamefont {C.}~\bibnamefont {Lee}}, \bibinfo {author}
  {\bibfnamefont {L.}~\bibnamefont {Wang}}, \bibinfo {author} {\bibfnamefont
  {S.}~\bibnamefont {Sorgenfrei}}, \bibinfo {author} {\bibfnamefont
  {K.}~\bibnamefont {Watanabe}}, \bibinfo {author} {\bibfnamefont
  {T.}~\bibnamefont {Taniguchi}}, \bibinfo {author} {\bibfnamefont
  {P.}~\bibnamefont {Kim}}, \bibinfo {author} {\bibfnamefont {K.~L.}\
  \bibnamefont {Shepard}}, \ and\ \bibinfo {author} {\bibfnamefont
  {J.}~\bibnamefont {Hone}},\ }\href {\doibase 10.1038/nnano.2010.172}
  {\bibfield  {journal} {\bibinfo  {journal} {Nature nanotechnology}\ }\textbf
  {\bibinfo {volume} {5}},\ \bibinfo {pages} {722} (\bibinfo {year}
  {2010})}\BibitemShut {NoStop}%
\bibitem [{\citenamefont {Han}\ \emph {et~al.}(2014)\citenamefont {Han},
  \citenamefont {Kawakami}, \citenamefont {Gmitra},\ and\ \citenamefont
  {Fabian}}]{han14}%
  \BibitemOpen
  \bibfield  {author} {\bibinfo {author} {\bibfnamefont {W.}~\bibnamefont
  {Han}}, \bibinfo {author} {\bibfnamefont {R.~K.}\ \bibnamefont {Kawakami}},
  \bibinfo {author} {\bibfnamefont {M.}~\bibnamefont {Gmitra}}, \ and\ \bibinfo
  {author} {\bibfnamefont {J.}~\bibnamefont {Fabian}},\ }\href {\doibase
  10.1038/nnano.2014.214} {\bibfield  {journal} {\bibinfo  {journal} {Nature
  nanotechnology}\ }\textbf {\bibinfo {volume} {9}},\ \bibinfo {pages} {794}
  (\bibinfo {year} {2014})}\BibitemShut {NoStop}%
\bibitem [{\citenamefont {Roche}\ \emph {et~al.}(2015)\citenamefont {Roche},
  \citenamefont {{\AA}kerman}, \citenamefont {Beschoten}, \citenamefont
  {Charlier}, \citenamefont {Chshiev}, \citenamefont {Dash}, \citenamefont
  {Dlubak}, \citenamefont {Fabian}, \citenamefont {Fert}, \citenamefont
  {Guimar{\~a}es}, \citenamefont {Guinea}, \citenamefont {Grigorieva},
  \citenamefont {Sch\"onenberger}, \citenamefont {Seneor}, \citenamefont
  {Stampfer}, \citenamefont {Valenzuela}, \citenamefont {Waintal},\ and\
  \citenamefont {\mbox{van Wees}}}]{roche15}%
  \BibitemOpen
  \bibfield  {author} {\bibinfo {author} {\bibfnamefont {S.}~\bibnamefont
  {Roche}}, \bibinfo {author} {\bibfnamefont {J.}~\bibnamefont {{\AA}kerman}},
  \bibinfo {author} {\bibfnamefont {B.}~\bibnamefont {Beschoten}}, \bibinfo
  {author} {\bibfnamefont {J.-C.}\ \bibnamefont {Charlier}}, \bibinfo {author}
  {\bibfnamefont {M.}~\bibnamefont {Chshiev}}, \bibinfo {author} {\bibfnamefont
  {S.~P.}\ \bibnamefont {Dash}}, \bibinfo {author} {\bibfnamefont
  {B.}~\bibnamefont {Dlubak}}, \bibinfo {author} {\bibfnamefont
  {J.}~\bibnamefont {Fabian}}, \bibinfo {author} {\bibfnamefont
  {A.}~\bibnamefont {Fert}}, \bibinfo {author} {\bibfnamefont {M.}~\bibnamefont
  {Guimar{\~a}es}}, \bibinfo {author} {\bibfnamefont {F.}~\bibnamefont
  {Guinea}}, \bibinfo {author} {\bibfnamefont {I.}~\bibnamefont {Grigorieva}},
  \bibinfo {author} {\bibfnamefont {C.}~\bibnamefont {Sch\"onenberger}},
  \bibinfo {author} {\bibfnamefont {P.}~\bibnamefont {Seneor}}, \bibinfo
  {author} {\bibfnamefont {C.}~\bibnamefont {Stampfer}}, \bibinfo {author}
  {\bibfnamefont {S.~O.}\ \bibnamefont {Valenzuela}}, \bibinfo {author}
  {\bibfnamefont {X.}~\bibnamefont {Waintal}}, \ and\ \bibinfo {author}
  {\bibfnamefont {B.}~\bibnamefont {\mbox{van Wees}}},\ }\href {\doibase
  10.1088/2053-1583/2/3/030202} {\bibfield  {journal} {\bibinfo  {journal} {2D
  Materials}\ }\textbf {\bibinfo {volume} {2}},\ \bibinfo {pages} {030202}
  (\bibinfo {year} {2015})}\BibitemShut {NoStop}%
\bibitem [{\citenamefont {Lafont}\ \emph {et~al.}(2015)\citenamefont {Lafont},
  \citenamefont {Ribeiro-Palau}, \citenamefont {Kazazis}, \citenamefont
  {Michon}, \citenamefont {Couturaud}, \citenamefont {Consejo}, \citenamefont
  {Chassagne}, \citenamefont {Zielinski}, \citenamefont {Portail},
  \citenamefont {Jouault}, \citenamefont {Schopfer},\ and\ \citenamefont
  {Poirier}}]{lafon15}%
  \BibitemOpen
  \bibfield  {author} {\bibinfo {author} {\bibfnamefont {F.}~\bibnamefont
  {Lafont}}, \bibinfo {author} {\bibfnamefont {R.}~\bibnamefont
  {Ribeiro-Palau}}, \bibinfo {author} {\bibfnamefont {D.}~\bibnamefont
  {Kazazis}}, \bibinfo {author} {\bibfnamefont {A.}~\bibnamefont {Michon}},
  \bibinfo {author} {\bibfnamefont {O.}~\bibnamefont {Couturaud}}, \bibinfo
  {author} {\bibfnamefont {C.}~\bibnamefont {Consejo}}, \bibinfo {author}
  {\bibfnamefont {T.}~\bibnamefont {Chassagne}}, \bibinfo {author}
  {\bibfnamefont {M.}~\bibnamefont {Zielinski}}, \bibinfo {author}
  {\bibfnamefont {M.}~\bibnamefont {Portail}}, \bibinfo {author} {\bibfnamefont
  {B.}~\bibnamefont {Jouault}}, \bibinfo {author} {\bibfnamefont
  {F.}~\bibnamefont {Schopfer}}, \ and\ \bibinfo {author} {\bibfnamefont
  {W.}~\bibnamefont {Poirier}},\ }\href {\doibase 10.1038/ncomms7806}
  {\bibfield  {journal} {\bibinfo  {journal} {Nature communications}\ }\textbf
  {\bibinfo {volume} {6}},\ \bibinfo {pages} {6806} (\bibinfo {year}
  {2015})}\BibitemShut {NoStop}%
\bibitem [{\citenamefont {Zheng}\ and\ \citenamefont {Ando}(2002)}]{zheng02c}%
  \BibitemOpen
  \bibfield  {author} {\bibinfo {author} {\bibfnamefont {Y.}~\bibnamefont
  {Zheng}}\ and\ \bibinfo {author} {\bibfnamefont {T.}~\bibnamefont {Ando}},\
  }\href {\doibase 10.1103/PhysRevB.65.245420} {\bibfield  {journal} {\bibinfo
  {journal} {Phys. Rev. B}\ }\textbf {\bibinfo {volume} {65}},\ \bibinfo
  {pages} {245420} (\bibinfo {year} {2002})}\BibitemShut {NoStop}%
\bibitem [{\citenamefont {McClure}(1956)}]{mcclu56}%
  \BibitemOpen
  \bibfield  {author} {\bibinfo {author} {\bibfnamefont {J.~W.}\ \bibnamefont
  {McClure}},\ }\href {\doibase 10.1103/PhysRev.104.666} {\bibfield  {journal}
  {\bibinfo  {journal} {Physical Review}\ }\textbf {\bibinfo {volume} {104}},\
  \bibinfo {pages} {666} (\bibinfo {year} {1956})}\BibitemShut {NoStop}%
\bibitem [{\citenamefont {Delplace}\ and\ \citenamefont
  {Montambaux}(2010)}]{delpl10}%
  \BibitemOpen
  \bibfield  {author} {\bibinfo {author} {\bibfnamefont {P.}~\bibnamefont
  {Delplace}}\ and\ \bibinfo {author} {\bibfnamefont {G.}~\bibnamefont
  {Montambaux}},\ }\href {\doibase 10.1103/PhysRevB.82.205412} {\bibfield
  {journal} {\bibinfo  {journal} {Phys. Rev. B}\ }\textbf {\bibinfo {volume}
  {82}},\ \bibinfo {pages} {205412} (\bibinfo {year} {2010})}\BibitemShut
  {NoStop}%
\bibitem [{\citenamefont {Aharonov}\ and\ \citenamefont
  {Bohm}(1959)}]{aharo59}%
  \BibitemOpen
  \bibfield  {author} {\bibinfo {author} {\bibfnamefont {Y.}~\bibnamefont
  {Aharonov}}\ and\ \bibinfo {author} {\bibfnamefont {D.}~\bibnamefont
  {Bohm}},\ }\href {\doibase 10.1103/PhysRev.115.485} {\bibfield  {journal}
  {\bibinfo  {journal} {Physical Review}\ }\textbf {\bibinfo {volume} {115}},\
  \bibinfo {pages} {485} (\bibinfo {year} {1959})}\BibitemShut {NoStop}%
\bibitem [{\citenamefont {Polizzi}(2009)}]{poliz09}%
  \BibitemOpen
  \bibfield  {author} {\bibinfo {author} {\bibfnamefont {E.}~\bibnamefont
  {Polizzi}},\ }\href {\doibase 10.1103/PhysRevB.79.115112} {\bibfield
  {journal} {\bibinfo  {journal} {Phys. Rev. B}\ }\textbf {\bibinfo {volume}
  {79}},\ \bibinfo {pages} {115112} (\bibinfo {year} {2009})}\BibitemShut
  {NoStop}%
\bibitem [{\citenamefont {Chen}\ \emph {et~al.}(2007)\citenamefont {Chen},
  \citenamefont {Lin}, \citenamefont {Rooks},\ and\ \citenamefont
  {Avouris}}]{chen07b}%
  \BibitemOpen
  \bibfield  {author} {\bibinfo {author} {\bibfnamefont {Z.}~\bibnamefont
  {Chen}}, \bibinfo {author} {\bibfnamefont {Y.-M.}\ \bibnamefont {Lin}},
  \bibinfo {author} {\bibfnamefont {M.~J.}\ \bibnamefont {Rooks}}, \ and\
  \bibinfo {author} {\bibfnamefont {P.}~\bibnamefont {Avouris}},\ }\href
  {\doibase 10.1016/j.physe.2007.06.020} {\bibfield  {journal} {\bibinfo
  {journal} {Physica {E}: {L}ow-dimensional {S}ystems and {N}anostructures}\
  }\textbf {\bibinfo {volume} {40}},\ \bibinfo {pages} {228} (\bibinfo {year}
  {2007})}\BibitemShut {NoStop}%
\bibitem [{\citenamefont {Hill}\ \emph {et~al.}(2006)\citenamefont {Hill},
  \citenamefont {Geim}, \citenamefont {Novoselov}, \citenamefont {Schedin},\
  and\ \citenamefont {Blake}}]{hill06}%
  \BibitemOpen
  \bibfield  {author} {\bibinfo {author} {\bibfnamefont {E.~W.}\ \bibnamefont
  {Hill}}, \bibinfo {author} {\bibfnamefont {A.~K.}\ \bibnamefont {Geim}},
  \bibinfo {author} {\bibfnamefont {S.~K.}\ \bibnamefont {Novoselov}}, \bibinfo
  {author} {\bibfnamefont {F.}~\bibnamefont {Schedin}}, \ and\ \bibinfo
  {author} {\bibfnamefont {P.}~\bibnamefont {Blake}},\ }\href {\doibase
  10.1109/TMAG.2006.878852} {\bibfield  {journal} {\bibinfo  {journal} {IEEE
  transactions on magnetics}\ }\textbf {\bibinfo {volume} {42}},\ \bibinfo
  {pages} {2694} (\bibinfo {year} {2006})}\BibitemShut {NoStop}%
\bibitem [{\citenamefont {Campos}\ \emph {et~al.}(2009)\citenamefont {Campos},
  \citenamefont {Manfrinato}, \citenamefont {Sanchez-Yamagishi}, \citenamefont
  {Kong},\ and\ \citenamefont {Jarillo-Herrero}}]{campo09}%
  \BibitemOpen
  \bibfield  {author} {\bibinfo {author} {\bibfnamefont {L.~C.}\ \bibnamefont
  {Campos}}, \bibinfo {author} {\bibfnamefont {V.~R.}\ \bibnamefont
  {Manfrinato}}, \bibinfo {author} {\bibfnamefont {J.~D.}\ \bibnamefont
  {Sanchez-Yamagishi}}, \bibinfo {author} {\bibfnamefont {J.}~\bibnamefont
  {Kong}}, \ and\ \bibinfo {author} {\bibfnamefont {P.}~\bibnamefont
  {Jarillo-Herrero}},\ }\href {\doibase 10.1021/nl900811r} {\bibfield
  {journal} {\bibinfo  {journal} {Nano Letters}\ }\textbf {\bibinfo {volume}
  {9}},\ \bibinfo {pages} {2600} (\bibinfo {year} {2009})}\BibitemShut
  {NoStop}%
\end{thebibliography}


\end{document}